\documentclass[conference]{IEEEtran}

\usepackage[utf8]{inputenc}
\usepackage[T1]{fontenc}
\usepackage[english]{babel}
\usepackage{microtype}
\usepackage{cite}
\usepackage{graphicx}
\graphicspath{{images/}}
\usepackage{booktabs}
\usepackage{tabularx}
\usepackage{array}
\usepackage{enumitem}
\usepackage{xcolor}
\usepackage{url}
\usepackage{stfloats}
\usepackage{balance}
\usepackage[hidelinks]{hyperref}
\hypersetup{
  pdftitle={The Data Hospital: A Workflow-Based Concept for Explainable Research Data Quality Assistance},
  pdfauthor={Lennard Scheurer, Robert Porzel, Vinicius Carrillo Beber, Rainer Malaka},
  pdfsubject={Concept paper on human-in-the-loop research data quality assistance},
  pdfkeywords={research data quality, human-in-the-loop, explainable assistance, provenance, reproducibility, research data management}
}

\newcommand{\datahospital}{\textit{Data Hospital}}
\newcommand{\drdata}{\textit{Dr. Data}}
\newcolumntype{Y}{>{\raggedright\arraybackslash}X}
\newcolumntype{L}[1]{>{\raggedright\arraybackslash}p{#1}}
\newcommand{\smalltable}{\footnotesize\setlength{\tabcolsep}{4pt}\renewcommand{\arraystretch}{1.12}}
\newcommand{\columnsmalltable}{\scriptsize\setlength{\tabcolsep}{3pt}\renewcommand{\arraystretch}{1.08}}

\title{The Data Hospital: A Workflow-Based Concept for Explainable Research Data Quality Assistance}

\author{
\IEEEauthorblockN{Lennard Scheurer}
\IEEEauthorblockA{
University of Bremen\\
Bremen, Germany\\
scheurer@uni-bremen.de
}
\and
\IEEEauthorblockN{Robert Porzel}
\IEEEauthorblockA{
University of Bremen\\
Bremen, Germany\\
porzel@tzi.de
}
\and
\IEEEauthorblockN{Vinicius Carrillo Beber}
\IEEEauthorblockA{
Fraunhofer IFAM\\
Bremen, Germany\\
vinicius.carrillo.beber@ifam.fraunhofer.de
}
\and
\IEEEauthorblockN{Rainer Malaka}
\IEEEauthorblockA{
University of Bremen\\
Bremen, Germany\\
malaka@tzi.de
}
}

\begin{document}

\maketitle
\pagestyle{plain}
\thispagestyle{plain}

\begin{abstract}
Research data quality is multidimensional and purpose-dependent: it emerges from the interplay of data, intended use, contextual knowledge, documentation, intervention decisions, and traceability. This concept paper presents the \datahospital{}, a human-in-the-loop control and interaction model for research data quality. Using a hospital metaphor, datasets are admitted, contextualized, assessed, reviewed in specialized stations, modified only through approved interventions, validated, documented, and made replayable where interventions are sufficiently specified. The concept combines deterministic profiling and inspectable evidence with optional evidence-bound explanation by \drdata{} and explicit user decisions. Preserved Raw Data and controlled working states separate observation from intervention. The contribution is not a new cleaning or imputation algorithm, but a ten-stage workflow that makes assessability, uncertainty, intervention authority, provenance, and process reproducibility visible. The prototype is an implementation-backed demonstrator rather than a released research artifact and illustrates selected parts of the concept through representational standardization, imputation, Patient File documentation, and replay. The paper concludes with a staged agenda for subsequent technical and user-centered evaluation.
\end{abstract}


\section{Introduction}

Researchers work with tabular data whose quality must be assessed and improved before analysis, archiving, model training, or publication. Typical problems include missing values, implicit missing-code markers, inconsistent categories, non-uniform units, formatting problems, duplicates, unclear metadata, or missing provenance. At the same time, data quality is context-dependent: an outlier may be a measurement error or a scientifically relevant extreme value; a value such as \texttt{NA} may denote a missing value, a category, or a domain-specific code. Data quality work is therefore not merely the detection of errors, but a process of interpretation, decision-making, intervention, validation, and documentation. This perspective connects to data quality research that treats quality as use-dependent and multidimensional and that distinguishes quality dimensions from their concrete assessment \cite{wang1996,pipino2002,batini2006}.

One motivation for the \datahospital{} emerged from work on missing data and imputation. Comparing imputation methods can quantify how well missing values are reconstructed under controlled conditions, but an imputation score alone cannot determine whether a research dataset is fit for a particular use. A dataset can be complete but difficult to interpret, representationally consistent but outdated, or technically clean while still lacking the context required for a scientifically defensible decision. The same observation applies beyond imputation: research data quality is multidimensional and purpose-dependent, and some of its most consequential steps require explicit human judgment.

Existing tools support many of these activities, including transformation, rule checking, repair, error detection, interactive curation, operation histories, provenance, and research-object packaging \cite{kandel2011,dallachiesa2013,rekatsinas2017,stonebraker2013,krishnan2016activeclean,mahdavi2019raha,openrefine_running,openrefine_exporting,rocrate2026}. They do not, however, necessarily foreground the same control questions in one researcher-facing interaction model: what can be assessed from the available evidence, where contextual judgment is required, who authorizes a consequential intervention, which data state changes, and how the decision remains inspectable. The \datahospital{} addresses this design opportunity through a hospital metaphor. A dataset is admitted, examined, assessed, modified through controlled interventions, monitored, and prepared for discharge, while its Patient File preserves the context and history needed to understand what happened and why.

\subsection{Problem Statement and Research Question}

Research data quality work involves more than detecting and correcting technical problems.
Quality assessments may depend on incomplete evidence and contextual knowledge, while interventions can alter the data and therefore require explicit decisions, validation, and documentation.
The resulting challenge is to organize these activities so that the basis of a quality claim, the authority to change the data, and the resulting data state remain visible throughout the process.

Accordingly, this concept paper is guided by the following research question:

\begin{quote}
How can a human-in-the-loop control workflow structure research data quality assessment and intervention such that evidence, human decisions, and resulting data-state changes remain understandable, controllable, and traceable?
\end{quote}

The question deliberately concerns the surrounding control and interaction model rather than the performance of a particular cleaning, repair, or imputation method.
It therefore focuses on how heterogeneous data-quality activities can be embedded in explicit control points for assessment, decision-making, intervention, validation, documentation, and reproduction.

\subsection{Positioning and Scope of this Paper}

This paper presents the conceptual and methodological framing of the \datahospital{} together with a non-public executable prototype that illustrates selected instantiations of the concept. The prototype demonstrates the workflow from dataset admission and contextualization through assessment, review, controlled intervention, validation, documentation, replay, and finalization. Station-specific methodological depth and implementation maturity can differ; the paper therefore distinguishes the stable conceptual model from the current scope of the prototype.

The prototype serves as an implementation-backed demonstrator rather than as a released research artifact or as evidence of empirical effectiveness. Its role is to show how the proposed control model, roles, data states, and documentation mechanisms can be instantiated in an executable interface.

\subsection{Contribution of the Paper}

The paper makes five conceptual contributions:

\begin{description}[leftmargin=*,style=nextline]
\item[C1] It connects multidimensional data quality and fitness for use to a station-based \datahospital{} model while making the evidential boundary and assessability of individual dimensions explicit.
\item[C2] It defines a general workflow (Contextualize, Detect, Assess, Justify, Explain, Decide, Intervene, Validate, Document, Reproduce) that structures data quality work beyond a single cleaning operation.
\item[C3] It separates deterministic analysis and intervention logic, optional evidence-bound LLM explanation, and explicit user control so that assistance does not become an LLM autopilot.
\item[C4] It illustrates the workflow through a non-public executable prototype comprising Patient Vitals, representation review, the Imputation Station, Patient File, Audit Log, validation, Final Report, and replay while preserving Raw Data and controlled data states.
\item[C5] It defines an evaluation agenda for the subsequent empirical assessment of the concept and its prototype, covering technical control, end-to-end usability, explanation, decision support, transparency, documentation, and replay.
\end{description}

\section{Background and Related Work}

This section summarizes the scientific and technical reference points from which the \datahospital{} is derived. It covers established data-quality concepts, interactive cleaning and repair, research data management, provenance, missing-data work, and LLM-supported assistance before identifying the research gap addressed by the artifact.

\subsection{Data Quality Dimensions and Fitness for Use}

A central starting point is the data quality model by Wang and Strong, which considers data quality from the perspective of data consumers and groups dimensions into intrinsic, contextual, representational, and accessibility data quality \cite{wang1996}. In addition, Pipino, Lee, and Wang show that data quality assessment must connect objective and subjective metrics in order to make quality statements practically useful \cite{pipino2002}. Batini and Scannapieco systematize data quality dimensions, assessment, and improvement methods and show that data quality has technical, organizational, and procedural components \cite{batini2006,batini2009}. This understanding is particularly relevant for research data because quality is not determined by technical correctness alone, but by \textit{fitness for use}. A dataset may be sufficient for an exploratory analysis but insufficient for a causal study.

The W3C Data Quality Vocabulary likewise distinguishes between abstract quality dimensions, metrics, quality measurements, and quality annotations \cite{w3c2016dqv}. This implies that a dimension such as accuracy or timeliness is not automatically measurable. It must be translated into a concrete measurement, a proxy, or an explicit statement of non-assessability. For research data, this distinction is central because many quality statements are not robust without domain knowledge, reference data, or usage context.

\subsection{Data Cleaning, Data Wrangling, and Data Repair}

Classical data cleaning and data wrangling systems address important subproblems. Potter's Wheel integrates transformation and discrepancy detection in an interactive environment \cite{raman2001}. Wrangler supports the interactive specification of data transformations and suggests transformations based on direct manipulation \cite{kandel2011}. OpenRefine supports interactive cleaning and transformation as well as persistent undo/redo history, exportable operation sequences, and project archives that retain prior states \cite{openrefine_running,openrefine_exporting}. Rule-based and repair-oriented systems such as NADEEF allow heterogeneous data quality rules to be specified \cite{dallachiesa2013,ebaid2013}. Probabilistic approaches such as HoloClean combine quality rules, statistical inference, and additional signals for data repair \cite{rekatsinas2017}. More recent systems extend this perspective: Data Tamer addresses end-to-end data curation with human involvement \cite{stonebraker2013}, ActiveClean models cleaning as an iterative process for statistical modeling \cite{krishnan2016activeclean}, Raha reduces the configuration effort in error detection \cite{mahdavi2019raha}, Baran addresses error correction with context representation and transfer learning \cite{mahdavi2020baran}, and KATARA combines knowledge bases and crowdsourcing for semantic data cleaning \cite{chu2015katara}. Surveys further show that error detection and data cleaning are technically broad research areas, but often focus strongly on individual error classes, rules, or repair models \cite{rahm2000,abedjan2016,ilyas2019}.

These systems show that transformation, rule checking, and repair can be powerfully supported in automated or semi-automated ways. For research data, however, deciding whether an observable pattern is actually a data quality problem often depends on measurement context, data collection logic, the codebook, later use, and documentation obligations. At this point, pure operational logic is not sufficient.

\begin{table*}[t]
\centering
\caption{Positioning of existing system directions in relation to the Data Hospital approach}
\smalltable
\begin{tabularx}{\textwidth}{L{0.18\textwidth}Y Y Y}
\toprule
\textbf{System direction} & \textbf{Central focus} & \textbf{Limitation for research data quality} & \textbf{Differentiation of the Data Hospital} \\
\midrule
OpenRefine / Wrangler & Interactive cleaning, transformation, exploration, and operation history. & Transformation provenance is supported, but research purpose, evidential limits, decision justification, and release context are not the central organizing model. & Connects operation history with assessability, review decisions, Patient File, and research-data export context. \\
ActiveClean / Data Tamer & Iterative cleaning for modeling or end-to-end data curation. & Oriented more toward model quality, integration, or curation pipelines than toward research data documentation. & Transfers the process perspective to auditable quality decisions and export readiness. \\
Raha / Baran / KATARA & Error detection, error correction, and semantic context use. & Powerful detection/correction, but no complete intervention history for research data. & Turns suggestions into review cases and links them to evidence, decision, and provenance. \\
NADEEF & Rule-based detection and repair of data errors. & Requires rules and is primarily oriented toward technical quality violations. & Treats rules as review suggestions and documents human decisions. \\
HoloClean & Probabilistic data repair with rules and statistical inference. & Repair can be powerful, but scientific justification and context checking remain separate tasks. & Understands repair not as an endpoint, but as an intervention that requires documentation. \\
FAIR assessment / RO-Crate & FAIR indicators, metadata, packaging, provenance, and reuse. & Primarily assess or package research objects rather than guide interactive quality interventions. & Uses metadata and provenance during assessment and intervention, while remaining compatible with downstream packaging. \\
\datahospital{} & Human-centered workflow from contextualization and assessment to decision, intervention, validation, documentation, and replay. & Not a replacement for specialized statistics or cleaning tools. & Integrates subtasks into a controllable research data quality process. \\
\bottomrule
\end{tabularx}
\end{table*}

\subsection{Human-in-the-Loop and Interactive Data Work}

Data quality work is highly interactive. Users must interpret indications, validate rules, accept or reject suggestions, and assess uncertainty. HCI research on interactive data analysis and visual transformation shows that direct manipulation, feedback, and iterative exploration are central design principles \cite{kandel2011,heer2012}. Interactive cleaning systems such as ActiveClean and KATARA further show that human decisions and examples can be not only interface actions, but part of the data quality logic itself \cite{krishnan2016activeclean,chu2015katara}. For assistance systems concerned with data quality, this implies that automation should not replace human control. Instead, the system should prepare, explain, and document decisions. For AI-supported assistance, human-AI interaction principles are additionally relevant, for example making uncertainty visible, maintaining user control, and making errors correctable \cite{amershi2019}.

In this context, human-in-the-loop does not merely mean that users press a button at the end. What matters is where human interpretation is necessary: in interpreting outliers, assessing non-uniform units, deciding about imputation, and approving changes for publication or archiving.

\subsection{FAIR, Metadata, and Research Data Management}

The FAIR principles articulate that digital research objects should be findable, accessible, interoperable, and reusable \cite{wilkinson2016}. The FAIR Data Maturity Model further translates the principles into assessment indicators without treating FAIRness as identical to scientific correctness or general data quality \cite{rda2020fair}. For practical publication, metadata standards such as DataCite, Dublin Core Metadata Terms, and DCAT support identification, citation, general-purpose description, and interoperable dataset catalogs \cite{datacite2026,dcmi2020,w3c2024dcat}. RO-Crate complements these standards by packaging research data with structured metadata, contextual entities, software, and provenance \cite{rocrate2026}. These approaches show that quality work does not end with cleaning a table: codebooks, provenance, access information, licenses, and export packages affect whether a dataset can later be understood and reused.

For an assistance system, it follows that metadata must not be treated as a downstream formality. They must be visible during diagnosis and intervention because they provide context for quality decisions. A missing-code marker, for instance, can only be interpreted correctly if it is known whether it represents a genuinely missing value, a non-applicable observation, or a domain-specific code.

\subsection{Provenance, Auditability, and Reproducibility}

Data interventions become more scientifically traceable and defensible when their decisions remain inspectable. Provenance models such as W3C PROV describe how entities, activities, and agents are connected \cite{w3c2013prov}. Research on provenance in scientific workflows further emphasizes that provenance information supports reproducibility, reuse, and traceability of data-related processes \cite{davidson2008provenance,herschel2017provenance}. Applied to data quality, this means documenting which issue was detected, what evidence was available, which intervention was proposed, who made the decision, which data version was affected, and which outputs were created.

Without auditability, a methodological problem arises. Cleaned data may look technically better, but become scientifically less traceable if it is no longer clear which values were changed, imputed, or excluded. For research data, therefore, not only the final state is relevant, but also the intervention history. Auditability and reproducibility are related but not identical: an audit trail can document what happened, while replay provides an additional operational check by reconstructing documented interventions on a preserved copy of the Raw Data and comparing the resulting state. In addition, it is important to be able to trace steps that do not proceed deterministically, such as when imputing new values.

\subsection{Missing Data and Imputation}

Missing values are not only a completeness problem; they affect analysis, modeling, and interpretability. The missing-data literature distinguishes, among others, assumptions such as MCAR, MAR, and MNAR and shows that the choice of an imputation method depends on data collection logic, variable roles, missingness patterns, and the analysis goal \cite{rubin1976,schafer2002,little2019,vanbuuren2018}. Practical methods such as Multiple Imputation by Chained Equations (MICE) and random-forest-based imputation illustrate that imputation strategies differ in their assumptions and handling of variable types \cite{azur2011,stekhoven2012,vanbuuren2018}. Imputation creates synthesized or derived values. This shifts the quality question: not only the originally missing value, but also the generated replacement value, its uncertainty, its assumptions, and its documentation become part of data quality.

For the \datahospital{}, imputation is therefore a particularly suitable example station. It makes visible that data intervention does not only correct existing values, but can create new data states. These new data states require their own quality indicators, flags, reports, and decisions about whether they are suitable for analysis, archiving, or publication.

\subsection{LLM-Supported Assistance, Trust, and Limitations}

LLM-based assistance can support data quality work if it is used as an explanatory add-on component rather than as an automatic truth source. Current work on LLM-supported table processing, semantic profiling, and data cleaning shows concrete fields of application: LLMs can support table tasks, semantic profiling, and rule-based cleaning workflows \cite{lu2025tablellm,huang2024cocoonprofile,zhang2025cocoonclean}. Research on automation use, appropriate reliance, and human-AI interaction shows that automated recommendations can be over- or under-relied upon and that uncertainty, user control, and correctability should remain visible \cite{parasuraman1997,lee2004,amershi2019}.

For a \datahospital{}, this leads to a design principle: an LLM assistant must not assert unsupported facts about accuracy, timeliness, or domain-specific correctness when the necessary context is missing. Answers must be grounded in concrete issues, metrics, rules, data profiles, or user context. The LLM explains and supports, but does not make final intervention decisions.

\subsection{Research Gap}

Existing work provides powerful methods for data cleaning, transformation, repair, provenance, FAIR assessment, imputation, interactive data work, and increasingly LLM-supported assistance. These approaches, however, generally focus on particular tasks, methods, or stages of the data-quality process.

For research data, an additional process-level problem remains: assessments can depend on incomplete evidence and contextual judgment, while interventions can alter the dataset and therefore require explicit authority, validation, and traceability. Existing approaches do not necessarily make assessability, inspectable evidence, human decision authority, controlled data states, documentation, and replay part of one consistent researcher-facing control model across heterogeneous quality tasks.

The \datahospital{} addresses this gap through a common control structure within which specialized assessment, cleaning, imputation, and packaging methods can remain independent.

\section{Research Approach and Design Requirements}
\label{sec:research-approach}

This section translates the gap identified in Section~II into an artifact-oriented research approach, derived requirements, and design principles. Keeping this derivation separate from the artifact description makes clear which claims originate in prior work and which choices constitute the proposed design.

\subsection{Design Science Framing}
\label{sec:design-science}

Methodologically, the \datahospital{} is treated as a design science artifact. The goal is not only to describe a phenomenon, but to construct an artifact for a relevant practical problem and define how it can subsequently be evaluated. Design science in information systems emphasizes problem relevance, artifact construction, and evaluation \cite{hevner2004}. The Design Science Research Methodology describes an iterative logic of problem identification, objective definition, design and development, demonstration, evaluation, and communication \cite{peffers2007}.

For the \datahospital{}, requirements are derived from literature and the problem statement and operationalized in an evolving prototype. In DSRM terms, the present concept paper covers problem identification through Sections~I--II, objectives and requirements through this section, design and development through the concept and the prototype in Section~IV, and demonstration through the walkthrough. Evaluation is intentionally specified as subsequent work in Section~V; communication is served by the present concept paper. This positioning allows the workflow and its control logic to be made explicit without claiming that every envisioned station is equally mature or empirically validated.

\subsection{Derived Requirements}
\label{sec:requirements}

The background review is a focused narrative synthesis rather than a systematic literature review. It yields the artifact requirements in Table~\ref{tab:requirements}, which cites the principal literature basis for each requirement and assigns identifiers R1--R7 for later traceability into the artifact and research agenda. Taken together, the requirements define four recurring concerns. First, quality claims must expose their epistemic boundary. Direct measurement and proxy use describe the evidence basis, while contextual dependence describes whether an interpretation requires additional knowledge. If the available evidence and context do not support a defensible claim, the dimension remains not assessable. Second, consequential changes require review because technical signals cannot resolve all semantic or methodological questions. Third, state changes must remain traceable through preserved Raw Data, provenance, documentation, and replay. Fourth, the process must connect table-level work with metadata, reuse, and release requirements rather than treating those concerns as unrelated downstream tasks.

\begin{table*}[!t]
\centering
\caption{Requirements for a Data Hospital artifact}
\label{tab:requirements}
\smalltable
\begin{tabularx}{\textwidth}{L{0.25\textwidth}Y Y}
\toprule
\textbf{Reference point} & \textbf{Problem} & \textbf{Requirement for the Data Hospital} \\
\midrule
\textbf{R1} DQ dimensions \cite{wang1996,pipino2002,w3c2016dqv} & Not every dimension is directly measurable or defensibly interpretable. & Record the evidence basis, contextual dependence, and limits of assessability. \\
\textbf{R2} Cleaning / Wrangling / Repair \cite{kandel2011,dallachiesa2013,rekatsinas2017} & Individual operations do not solve the entire research data process. & Connect contextualization, detection, assessment, review, intervention, validation, documentation, and export as a workflow. \\
\textbf{R3} Human-in-the-loop \cite{amershi2019,lee2004} & Automation can misinterpret context. & Require or recommend human review decisions at critical points. \\
\textbf{R4} FAIR / RDM \cite{wilkinson2016,rda2020fair,rocrate2026} & Metadata and packaging affect reuse. & Treat codebook, provenance, license, and readiness as quality-relevant components. \\
\textbf{R5} Provenance \cite{w3c2013prov,davidson2008provenance,herschel2017provenance} & Cleaned data without history are methodologically difficult to verify. & Document decisions, data states, and outputs and support replay of documented interventions from preserved Raw Data. \\
\textbf{R6} Imputation \cite{rubin1976,little2019,vanbuuren2018} & Synthesized values create new quality and documentation questions. & Mark imputed values, document assumptions, and generate imputation reports. \\
\textbf{R7} LLM assistance \cite{parasuraman1997,lee2004,amershi2019} & Explanations can miscalibrate trust. & Provide evidence-bound explanations that communicate uncertainty. \\
\bottomrule
\end{tabularx}
\end{table*}

\subsection{Design Principles}
\label{sec:design-principles}

The requirements are condensed into nine design principles that guide the artifact. These principles form the bridge between the research basis and the concrete Data Hospital workflow described in Section~\ref{sec:data-hospital}. They also make the intended boundaries of automation explicit. DP1, DP3, and DP4 constrain what the system may claim and recommend; DP2 and DP7 preserve human control around risky interventions and LLM-supported explanation; DP5 and DP8 protect provenance and data-state integrity; and DP6 and DP9 connect metadata and detailed audit information with the practical interaction flow. The principles are therefore not additional workflow stages, but design constraints that apply across stages and stations.

\begin{table}[!t]
\centering
\caption{Design principles of the Data Hospital}
\label{tab:design-principles}
\columnsmalltable
\begin{tabularx}{\columnwidth}{L{0.38\columnwidth}Y}
\toprule
\textbf{Principle} & \textbf{Meaning for the artifact} \\
\midrule
DP1 Make assessability explicit & Show whether a dimension is directly measurable, proxy-based, context-dependent, or not assessable. \\
DP2 Review before risky intervention & Do not apply potentially risky transformations without a human decision. \\
DP3 Evidence before recommendation & Every recommendation requires traceable evidence. \\
DP4 Show uncertainty & Missing context is a result, not a system failure. \\
DP5 Secure provenance and replay & Every relevant decision must be auditable, and sufficiently documented interventions should be replayable/traceable from Raw Data. \\
DP6 Treat metadata as quality & Quality does not end at the table. \\
DP7 LLM as assistant (not authority) & \drdata{} explains but does not decide automatically. \\
DP8 Protect raw data & Raw Data remain unchanged; modifications emerge in controlled data states. \\
DP9 Separate compact use from audit details & Current Table and Intervention Details serve different purposes. \\
\bottomrule
\end{tabularx}
\end{table}

\section{The Data Hospital}
\label{sec:data-hospital}

\subsection{Guiding Idea and Hospital Metaphor}
\label{sec:metaphor}

The \datahospital{} uses a hospital metaphor to make an otherwise abstract data-quality process visible and navigable. A dataset is understood as a patient. It is admitted, examined, assessed, modified through controlled interventions, monitored, and finally prepared for discharge. The metaphor is not merely a visual theme: it structures roles, states, decision points, and documentation. At the same time, the metaphor has a deliberate limit. Data are not objectively ``sick'' or ``healthy'' in every context; whether an observed property is problematic depends on the intended use and on available domain knowledge.

The metaphor translates the process into concrete components: the Emergency Room supports admission and initial triage, Patient Vitals summarize quality assessments, specialized stations support interventions, the Patient File preserves context and history, \drdata{} supports explanation, and discharge leads to a Final Report and Export Package. Metaphorical names are paired with functional descriptions so that orientation does not depend on interpreting the metaphor. The central design claim is not that the system autonomously repairs data, but that useful automation is embedded in an explicit human-controlled process.

\subsection{Data Quality, Data Health, and Data Fitness}
\label{sec:data-health-fitness}

The conceptual starting point remains the multidimensional model of Wang and Strong \cite{wang1996}. The \datahospital{} treats intrinsic, contextual, representational, and accessibility data quality as categories of quality dimensions rather than reducing quality to a single technical score. A dimension describes \textit{what} aspect of quality is relevant, while metrics, rules, contextual statements, or manual assessments provide evidence about that dimension \cite{pipino2002,w3c2016dqv}.

To structure the artifact, this paper uses two additional system-level terms. \textit{Data Health} denotes the available measurements, indicators, evidence, and assessability statements used to characterize the dataset across quality dimensions. \textit{Data Fitness} denotes the purpose-specific selection and interpretation of quality dimensions and supporting evidence for a stated use or intervention goal. Data Fitness is therefore not a universal automatically computed ``fitness score.'' It is a purpose-specific view of the available quality evidence and may still require contextual or domain judgment. These terms do not replace established data-quality dimensions; they distinguish the overall observable quality state from purpose-specific fitness for use.

\begin{figure*}[!t]
\centering
\includegraphics[width=0.98\textwidth]{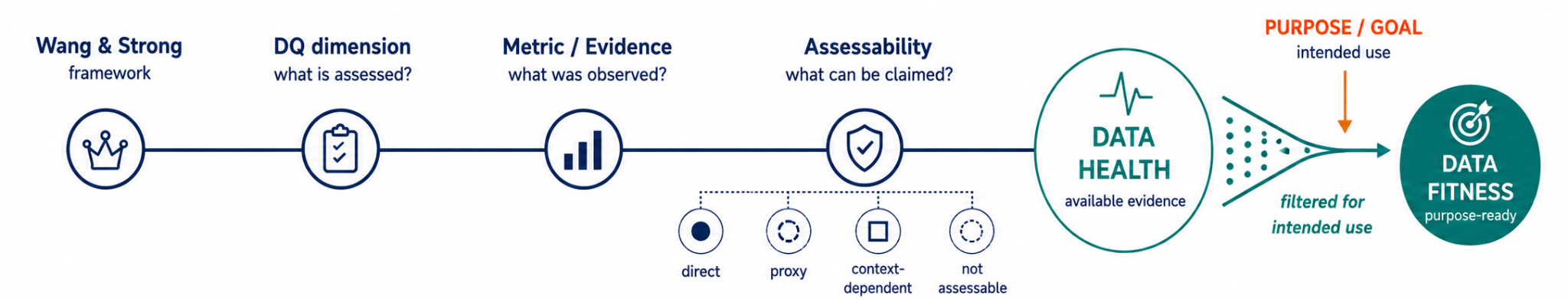}
\caption{Conceptual mapping used by Patient Vitals.}
\label{fig:quality-concept}
\end{figure*}

Figure~\ref{fig:quality-concept} summarizes this mapping. Assessability describes the strength and boundary of a claim, while the intended use filters available evidence into a purpose-specific fitness judgment. Workflow-specific readiness indicators such as Imputation Readiness or Publication Readiness remain supporting indicators rather than additional Wang--Strong dimensions. This implements R1, DP1, and DP4.

Table~\ref{tab:vitals-operationalization} makes the intended operationalization explicit. It does not claim that all Wang--Strong dimensions can be automated. Instead, it distinguishes evidence that is currently used in the prototype from dimensions that require additional context, reference data, or later implementation. Evidence basis and contextual dependence are conceptually separate: a proxy can also require contextual interpretation. The compact prototype interface summarizes these properties as display states, while a future operational model should retain the underlying evidence and context requirements. Three scope statuses remain distinct: \textit{not assessed in the prototype} means that a dimension belongs to the framework but is outside the current prototype assessment; \textit{not assessable} means that the available evidence or context does not support a defensible claim for the current case; and \textit{not implemented} refers to a planned function or station for which no executable prototype path is available.

\begin{table*}[!t]
\centering
\caption{Illustrative operationalization of the Wang and Strong dimensions in the current Data Hospital prototype.}
\label{tab:vitals-operationalization}
\scriptsize
\setlength{\tabcolsep}{3.5pt}
\renewcommand{\arraystretch}{1.10}
\begin{tabularx}{\textwidth}{L{0.19\textwidth}L{0.37\textwidth}Y}
\toprule
\textbf{Dimension} & \textbf{Evidence required or used in the prototype} & \textbf{Current prototype treatment} \\
\midrule
Accuracy & Reference or ground truth appropriate to the domain. & Not assessed in the prototype. \\
Objectivity & Methodological and domain context about how values were produced. & Not assessed in the prototype. \\
Believability & Source, provenance, and domain context supporting trustworthiness. & Not assessed in the prototype. \\
Reputation & Source, provenance, and external evidence about the standing or trustworthiness of the dataset or source. & Not assessed in the prototype; source and provenance metadata provide contextual evidence but do not establish reputation. \\
\addlinespace
Relevancy & Intended use and domain judgment about whether variables support the task. & Not assessed in the prototype. \\
Value-added & Purpose-specific judgment about whether the data add useful information. & Not assessed in the prototype. \\
Timeliness & A usable reference date or task-specific temporal requirement. & Context-dependent; explicitly not assessable when no reference date is available. \\
Completeness & Missing cells, missing-code markers, and related completeness indicators. & Implemented for supported completeness checks. \\
Appropriate amount & Purpose-specific judgment about whether the amount of data is suitable. & Not assessed in the prototype. \\
\addlinespace
Interpretability & Column semantics, labels, codebook information, and variable descriptions. & Implemented through semantic-information checks and metadata review. \\
Ease of understanding & Descriptions and explanatory metadata available to users. & Implemented through description and documentation checks. \\
Representational consistency & Category variants, missing-code conflicts, date, unit, type, and other representation issues. & Implemented for selected deterministic review families. \\
Concise representation & Redundant or duplicate records and avoidable repetition. & Implemented for selected structural checks. \\
\addlinespace
Accessibility & Access pathway and availability information. & Not assessed in the prototype. \\
Access security & Access-control and security information. & Not assessed in the prototype. \\
\bottomrule
\end{tabularx}
\end{table*}

\subsection{Overall Workflow}
\label{sec:workflow}

\begin{figure*}[!b]
\centering
\includegraphics[width=0.98\textwidth]{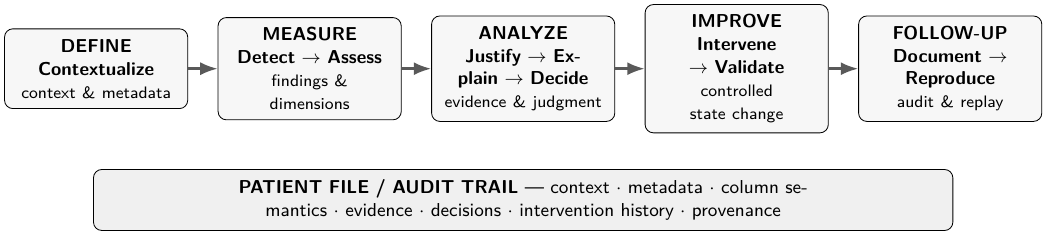}
\caption{The proposed Data Hospital control workflow, consisting of five phases from contextualization to documentation and reproduction, with the Patient File and Audit Trail spanning the process.}
\label{fig:workflow-concept}
\end{figure*}

The core contribution of the concept is a general control structure that is independent of any single station or intervention method. Figure~\ref{fig:workflow-concept} groups the ten stages into five process phases while the Patient File and Audit Trail span the workflow as a persistent record.

The stages have distinct responsibilities. \textit{Contextualize} captures intended use, metadata, column semantics, missing-code conventions, and known limitations. \textit{Detect} identifies technical patterns or possible quality issues. \textit{Assess} maps available evidence onto data-quality dimensions and makes non-assessability explicit. \textit{Justify} exposes the concrete evidence behind a finding. \textit{Explain} translates technical findings, alternatives, risks, and remaining uncertainty into an understandable form. \textit{Decide} keeps acceptance, rejection, editing, and uncertainty resolution under user control. \textit{Intervene} applies an approved transformation to a controlled working state rather than overwriting Raw Data. \textit{Validate} checks the result of that intervention. \textit{Document} records context, evidence, decisions, parameters, outputs, and provenance. Finally, \textit{Reproduce} replays sufficiently specified interventions to verify that the documented process can reconstruct the recorded state.

The five phases are a control sequence, not a strictly linear waterfall. Assessment can reveal missing context, explanation can expose insufficient evidence, and failed validation can return a case to assessment or decision. Documentation is continuous through the Patient File and Audit Trail rather than deferred until the end; reproduction is a cross-cutting capability invoked when the recorded process must be checked. The evidence and technical method can change from station to station, but the control points remain stable. This realizes R2 and provides the process-level basis for R3, DP2, and DP3.

\subsection{Deterministic Core, Dr. Data, and User Control}
\label{sec:roles}

The workflow separates three roles. First, the deterministic core is responsible for context and metadata contracts, profiling, quality metrics, evidence generation, confirmed transformations, validation, audit/provenance, and replay. Second, \drdata{} provides an optional explanation layer that can explain findings, translate technical context, answer questions, and discuss alternatives. It must not mutate data directly or replace missing evidence with invented certainty. Prepared evidence remains available without an LLM call. Third, users remain responsible for context confirmation and for decisions that require interpretation, including accepting, editing, rejecting, or postponing recommendations and choosing representation or imputation options.

\begin{figure*}[!t]
\centering
\includegraphics[width=0.92\textwidth]{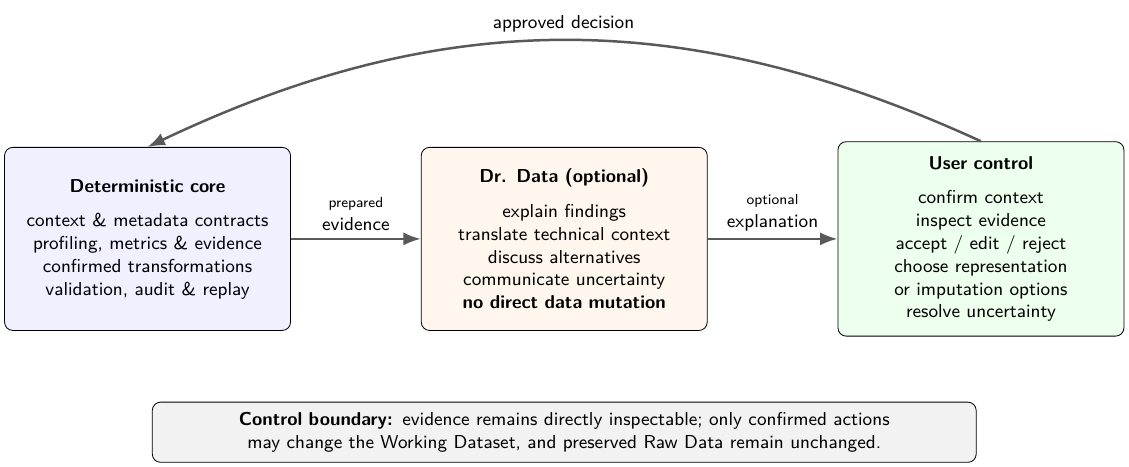}
\caption{Responsibility separation between the deterministic core, optional Dr. Data explanation, and explicit user control.}
\label{fig:role-separation}
\end{figure*}

Figure~\ref{fig:role-separation} summarizes this responsibility boundary. Evidence is available to the user independently of the LLM, the explanation layer remains optional, and only an explicit user-approved action is returned to the deterministic core for a state-changing intervention.

This separation prevents an LLM-autopilot design. A deterministic metric or rule can trigger a finding; \drdata{} can make the finding understandable; but the intervention path remains governed by evidence and explicit decisions. The role separation realizes R7 and DP7 and supports R3 and DP2. The more detailed control sequence remains the workflow defined in Section~\ref{sec:workflow}; the three roles describe responsibilities within that workflow rather than additional stages.

\subsection{Controlled Data States, Validation, and Replay}
\label{sec:states-validation-replay}

The \datahospital{} distinguishes between data states and views. Raw Data remain the preserved reference. Accepted state-changing interventions produce a controlled Working Dataset, and the Final Dataset is an approved export state. In compact form, the state path is therefore \textit{Raw Data $\rightarrow$ Working Dataset $\rightarrow$ Final Dataset}; views such as Raw View, Current Table, Intervention Details, and Export Preview expose these states without themselves creating new versions.

Validation and replay serve different purposes. Station-specific validation checks whether an approved intervention produced the expected result and whether unresolved cases remain visible. Replay instead starts from preserved Raw Data plus the Audit Trail, reconstructs sufficiently specified interventions, and compares the reconstructed state with the recorded reference state. A match therefore supports process reproducibility, not scientific correctness; a mismatch points back to the replay specification or audit history for inspection. At concept level, this state relation is deliberately kept simple; the operational replay behavior is demonstrated concretely by the prototype in Section~\ref{sec:documentation-export}. The state model realizes R5, DP5, DP8, and DP9.

\subsection{Core Components and Prototype}
\label{sec:components-stations}

The core components implement different responsibilities within the common workflow:

\begin{itemize}[leftmargin=*,nosep]
\item \textbf{Emergency Room:} dataset intake, uploaded supporting material, admission goal, and first triage.
\item \textbf{Identity Diffusion:} dataset context, column understanding, known limitations, and intervention planning.
\item \textbf{Patient Vitals:} multidimensional data-quality assessment, findings, warnings, and explicit assessability.
\item \textbf{Physical Therapy:} review and intervention for representational issues such as missing-code handling, category mappings, units, types, and dates.
\item \textbf{Imputation Station:} missing-value interventions that create and explicitly mark synthesized values.
\item \textbf{Patient File / Codebook:} canonical metadata, semantics, decisions, review status, and provenance.
\item \textbf{Audit Log and Replay:} event-level traceability and reconstruction of sufficiently documented interventions.
\item \textbf{Finalization and Export:} archive/publication readiness, Final Report, Final Dataset, and the Export Package for reuse and release.
\end{itemize}

The workflow constitutes the stable conceptual core of the \datahospital{}, while the non-public prototype provides one executable instantiation of selected elements. It serves to demonstrate how the proposed control structure can be realized in practice, rather than as a released or independently verifiable research artifact. Table~\ref{tab:concept-prototype} therefore distinguishes conceptual responsibilities from their current level of implementation, making clear which elements are operational, partial, or still planned. This separation allows individual stations to evolve without altering the underlying control model.

\begin{table*}[!t]
\centering
\caption{Concept-to-prototype matrix for the current Data Hospital prototype}
\label{tab:concept-prototype}
\smalltable
\begin{tabularx}{\textwidth}{L{0.20\textwidth}Y L{0.29\textwidth}}
\toprule
\textbf{Concept element} & \textbf{Responsibility in the concept} & \textbf{Current prototype status} \\
\midrule
Emergency Room & Admission, active-patient selection, supporting material, and goal selection. & Implemented for tabular intake and admission-path selection. \\
Identity Diffusion (Dataset Context and Semantics) & Contextualization, column understanding, limitations, and treatment planning. & Implemented for the demonstration workflow. \\
Patient Vitals & Multidimensional assessment with explicit assessability boundaries. & Implemented for selected metrics and dimensions; unsupported dimensions remain visibly not assessed or not assessable. \\
Physical Therapy (Representation Review) & Evidence-bound review and controlled representational interventions. & Implemented for selected missing-code, category, type, unit, and date review families. \\
Imputation Station & Consequential missing-value intervention with explicit configuration, marking, and validation. & Prototype implementation available; pre-imputation methodological depth can continue to mature. \\
\drdata{} & Optional explanation and decision support grounded in prepared evidence. & Explanation interface implemented; no direct data mutation. \\
Patient File / Audit / Replay & Persist context, decisions, state changes, and reconstruct sufficiently documented interventions. & Implemented for the demonstrated prototype path; replay depends on sufficiently specified history. \\
Finalization / Export & Assemble the final dataset and available documentation for reuse or release. & Core package implemented; unavailable optional artifacts remain explicitly marked unavailable. \\
Additional stations & Extend the same control structure to further quality interventions. & Not implemented unless explicitly exposed by the prototype; planned functions remain unavailable rather than simulated. \\
\bottomrule
\end{tabularx}
\end{table*}

\subsection{Prototype Demonstration: Admission and Contextualization}
\label{sec:walkthrough-intake}

\begin{figure*}[!tb]
\centering
\includegraphics[width=0.97\textwidth]{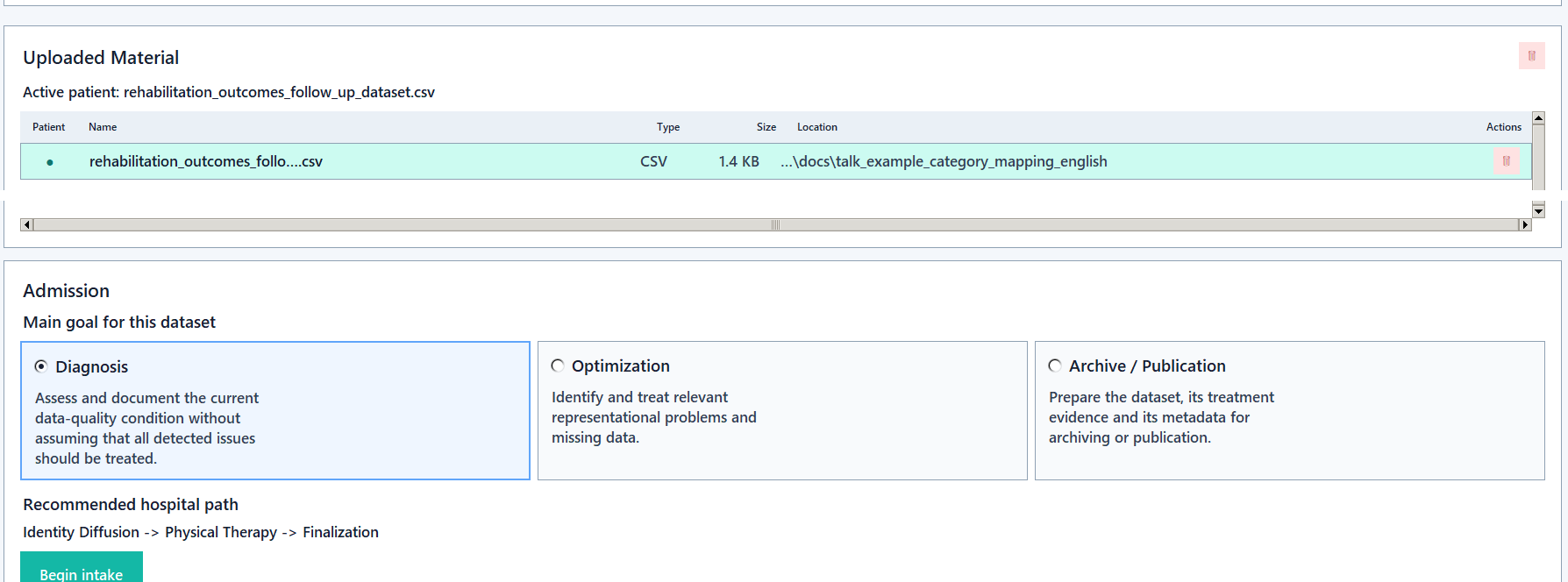}
\\[-0.2em]\footnotesize\textbf{(a)} Emergency Room: active patient and admission goal.
\par\vspace{0.35em}
\includegraphics[width=0.97\textwidth]{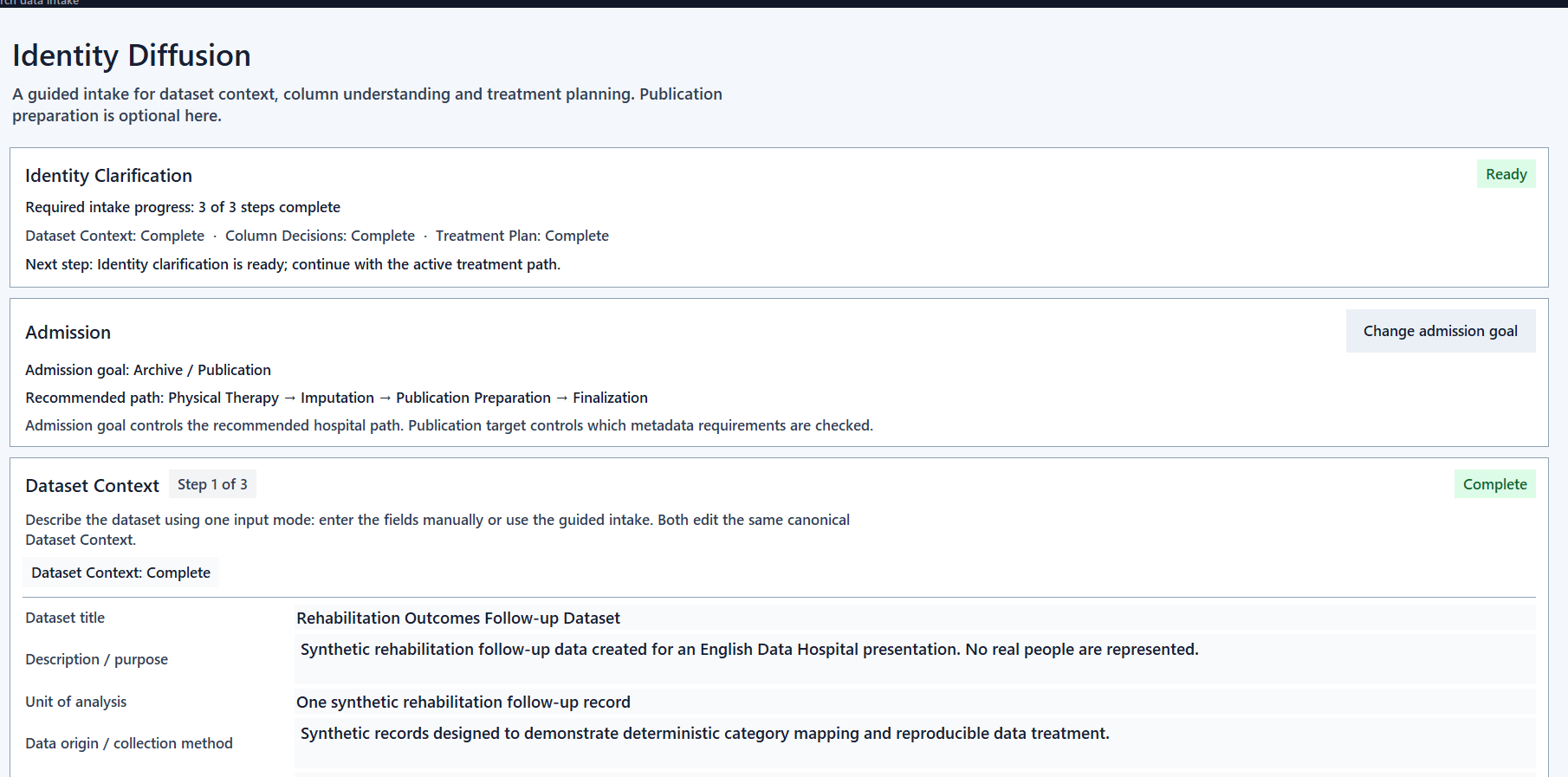}
\\[-0.2em]\footnotesize\textbf{(b)} Identity Diffusion (Dataset Context and Semantics) with dataset context and treatment planning.
\caption{Admission and contextualization in the current prototype.}
\label{fig:prototype-intake}
\end{figure*}

The prototype makes contextualization an explicit part of the workflow before quality interventions begin. Figure~\ref{fig:prototype-intake} shows the admission and contextualization steps at a readable interface scale. The Emergency Room makes the active dataset visible together with the admission goal, so that the patient entering the workflow is explicit. Identity Diffusion (Dataset Context and Semantics) then captures dataset purpose, unit of analysis, origin or collection method, known limitations, column decisions, and intervention planning in a canonical context record.

This ordering is important because later findings can depend on information that is not recoverable from the table alone. Missing-code conventions, intended use, and variable meaning are therefore treated as quality-relevant context rather than as optional documentation added only after cleaning.

\subsection{Prototype Demonstration: Patient Vitals}
\label{sec:walkthrough-vitals}

Patient Vitals translate multidimensional data quality into a visible assessment without claiming that every dimension is numerically measurable. The system can expose supported findings, completion states, and explicit non-assessability. Excluded or non-assessable metrics are not converted into artificial zero scores; instead, the boundary of what can be claimed from the available data and context remains visible.

\begin{figure*}[!tb]
\centering
\includegraphics[width=0.96\textwidth]{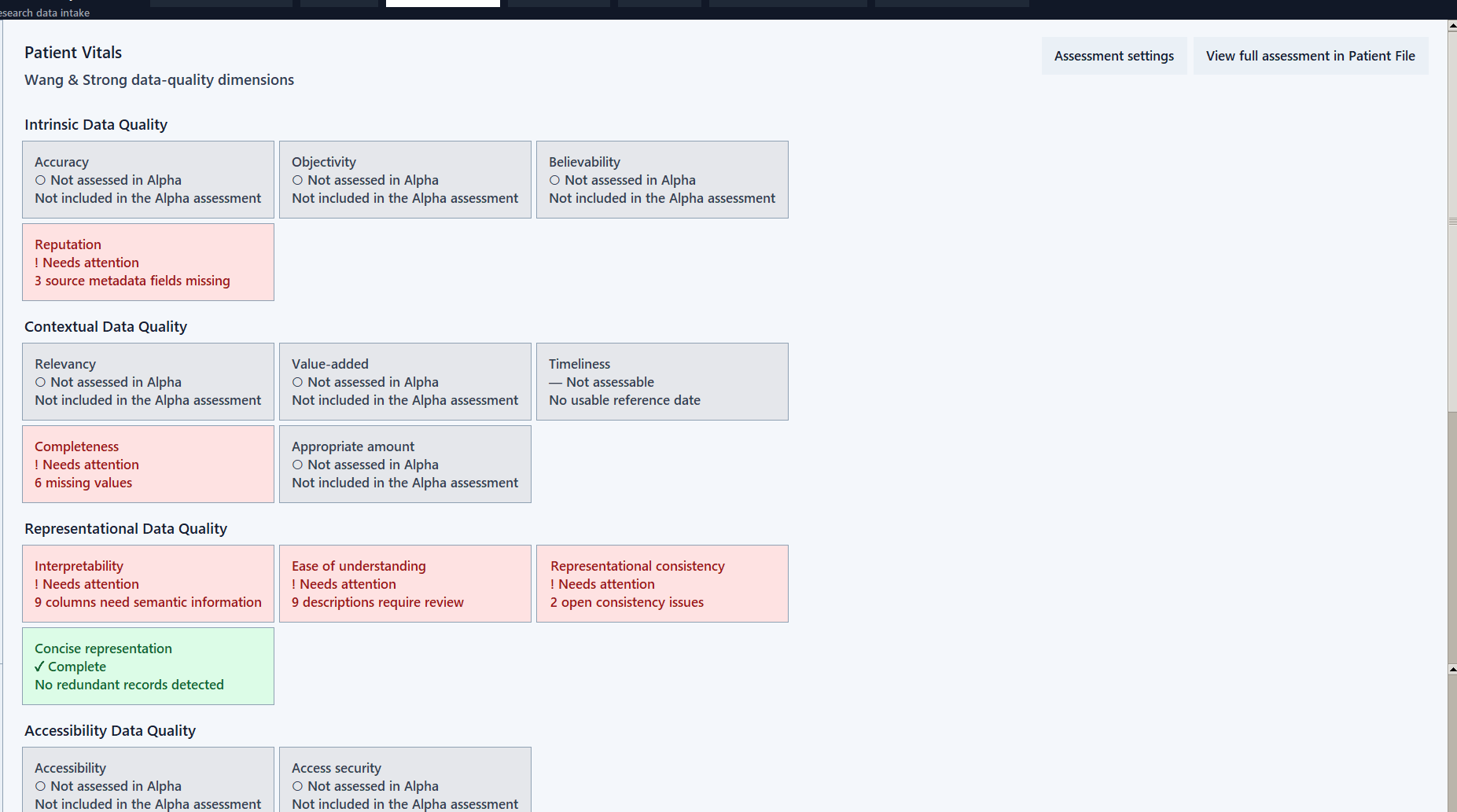}
\caption{Patient Vitals in the current prototype.}
\label{fig:prototype-vitals}
\end{figure*}

Figure~\ref{fig:prototype-vitals} shows the distinction at interaction level. In the synthetic example, some dimensions surface concrete warnings, concise representation is shown as complete, some dimensions are marked \textit{Not assessed in prototype}, and timeliness is marked \textit{Not assessable} because no usable reference date is available. The dashboard therefore acts as an assessment overview rather than as a single health score.

\subsection{Prototype Demonstration: Evidence, Explanation, and Decision}
\label{sec:walkthrough-representational}

Physical Therapy (Representation Review) implements representational interventions as inspectable review cases rather than as opaque cleaning commands. A detected problem is represented as a \textit{Quality Issue}, a proposed intervention as a \textit{Review Rule}, and affected cases as \textit{Review Cases}. The review view exposes evidence, the proposed mapping, expected impact, and unresolved values before a state change is accepted.

A category-standardization case illustrates the control logic. In the synthetic \texttt{consent} column, variants such as \texttt{yes}, \texttt{Yes}, \texttt{YES}, \texttt{no}, and \texttt{No} are evidence of inconsistent representation. They do not, however, establish that an ambiguous value such as \texttt{unknown} belongs to either canonical category. The deterministic rule can therefore propose mappings for high-confidence formatting variants while leaving ambiguous cases open.

\begin{figure*}[!tb]
\centering
\includegraphics[width=0.97\textwidth]{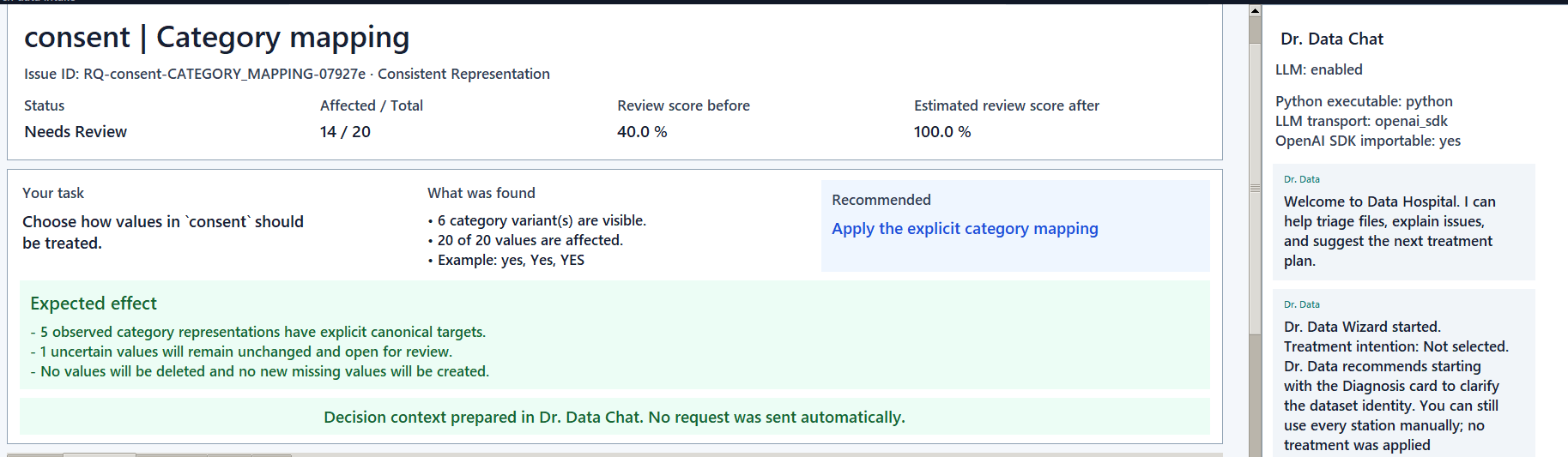}
\\[-0.2em]\footnotesize\textbf{(a)} Prepared evidence and decision context beside the optional \drdata{} panel.
\par\vspace{0.35em}
\includegraphics[width=0.80\textwidth]{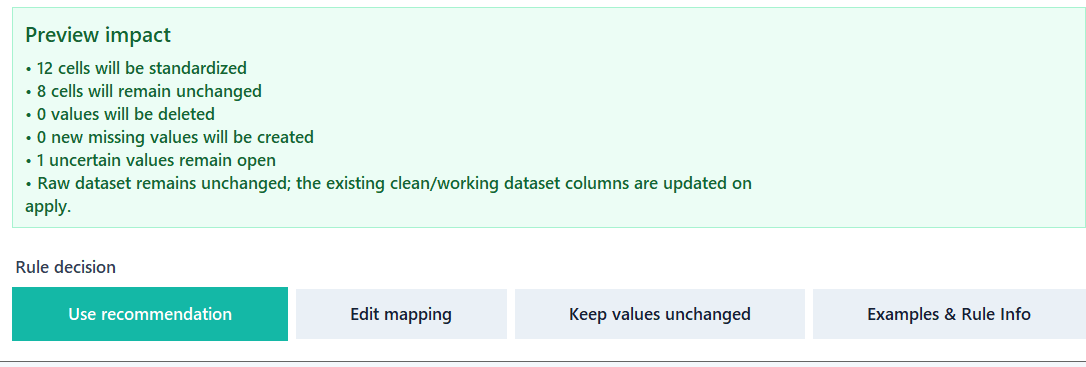}
\\[-0.2em]\footnotesize\textbf{(b)} Compact crop of the preview impact and explicit rule-decision controls.
\caption{Evidence-bound assistance and user control in the prototype.}
\label{fig:prototype-explanation-decision}
\end{figure*}

Figure~\ref{fig:prototype-explanation-decision}a shows the optional \drdata{} panel beside the same evidence and rule context. The interface explicitly states that decision context has been prepared and that no request is sent automatically. This matters because the explanation layer is subordinate to the evidence and does not initiate a state change. Figure~\ref{fig:prototype-explanation-decision}b focuses on the decision controls: the user can use the recommendation, edit the mapping, or keep values unchanged. The prototype therefore separates evidence, optional explanation, and explicit decision rather than collapsing them into an automatic repair.

\begin{figure*}[!tb]
\centering
\includegraphics[width=0.96\textwidth]{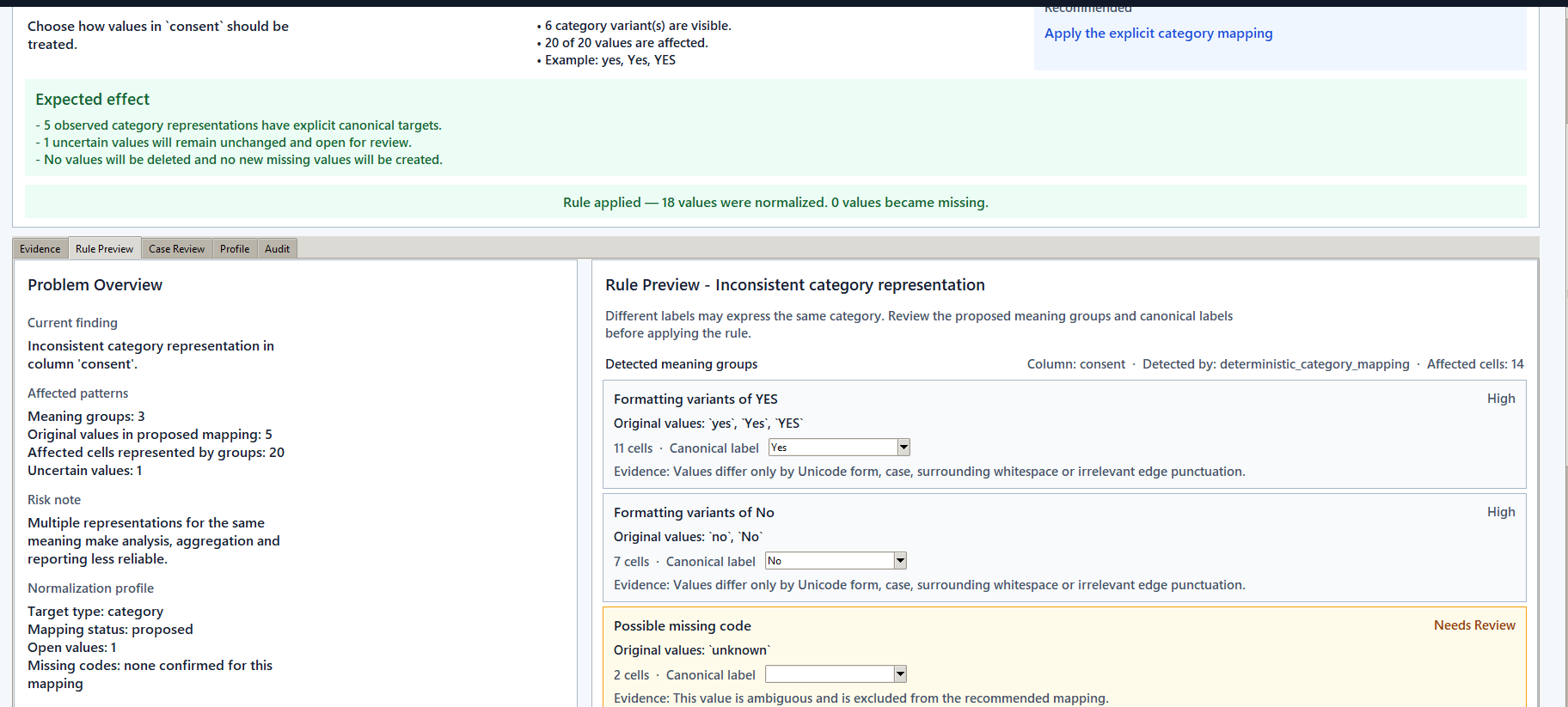}
\caption{Post-intervention view.}
\label{fig:prototype-after-treatment}
\end{figure*}

An approved mapping is applied only to the Working Dataset; Raw Data remain unchanged. Figure~\ref{fig:prototype-after-treatment} shows the post-intervention state. The result reports the applied normalization while the ambiguous value \texttt{unknown} remains marked as a possible missing code that needs review. The example illustrates the workflow with a comparatively simple intervention without making imputation the definition of the \datahospital{}.

\subsection{Imputation as a Consequential Intervention}
\label{sec:imputation-intervention}

\begin{figure*}[!t]
\centering
\includegraphics[width=0.96\textwidth]{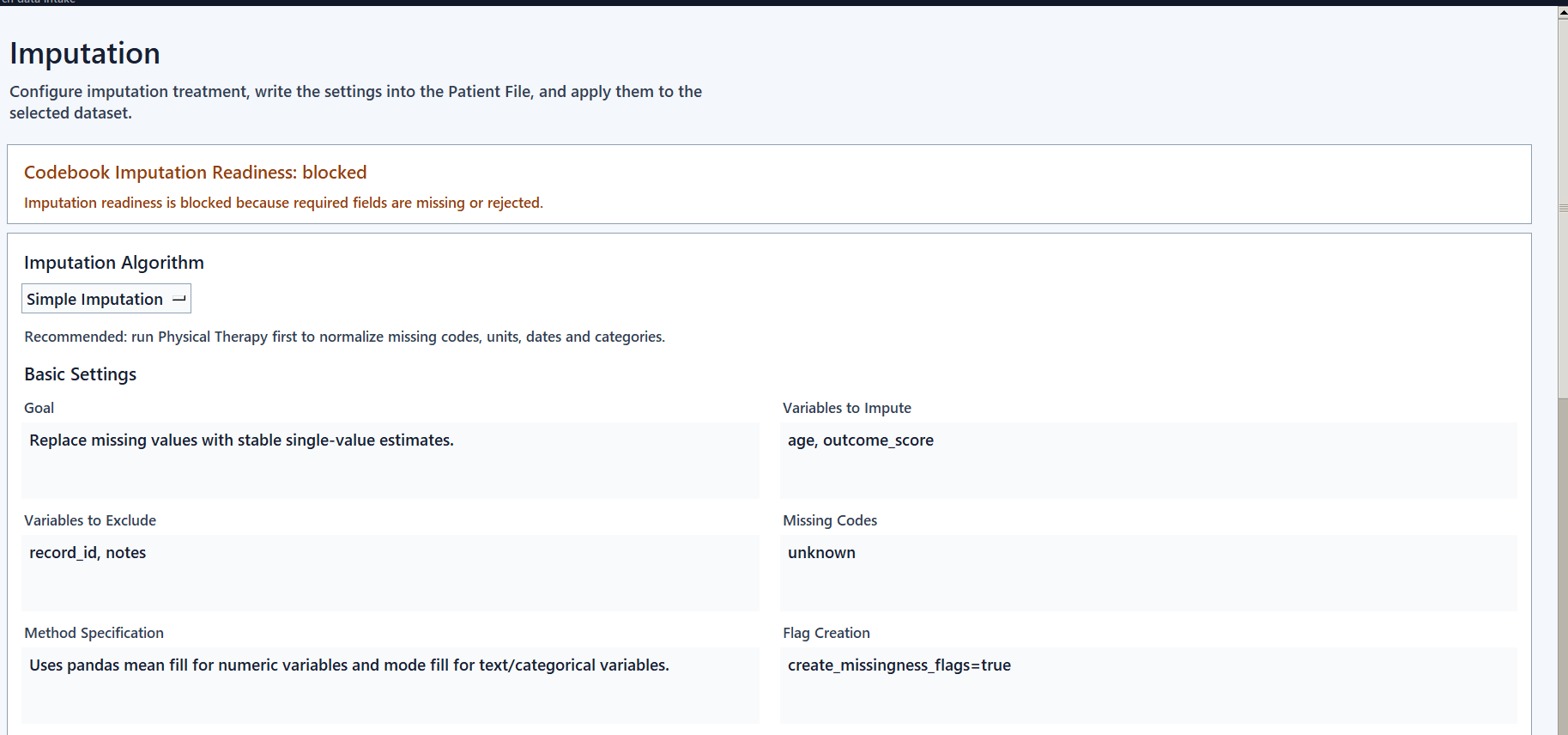}
\caption{Imputation readiness and method configuration before execution in the current prototype.}
\label{fig:prototype-imputation-config}
\end{figure*}

\begin{figure*}[!t]
\centering
\includegraphics[width=0.96\textwidth]{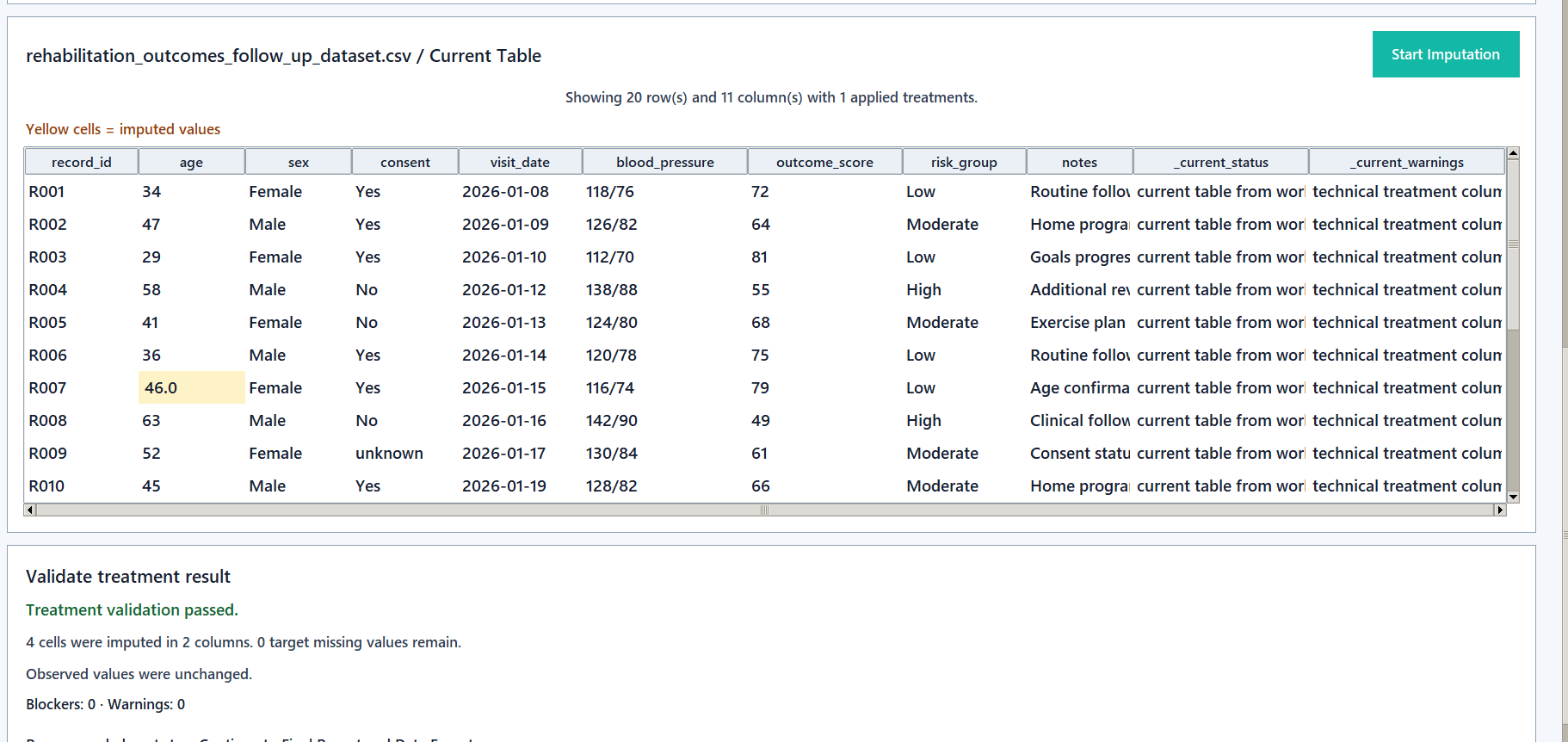}
\caption{Imputation result and validation in the current prototype.}
\label{fig:prototype-imputation-result}
\end{figure*}

The Imputation Station applies the same control structure to a more consequential intervention. Missing values are not treated as a purely technical replacement problem, but as a context-dependent decision whose assumptions and procedures must be considered against the missing-data literature \cite{rubin1976,schafer2002,little2019,vanbuuren2018}. Relevant considerations include missing-code markers, missingness patterns, possible structural missingness, variable roles, predictor availability, and the intended analysis or reuse goal. These determine whether imputation is appropriate at all or whether missing values should instead remain documented, excluded, or analyzed separately.

Because imputation creates synthesized or derived values, the Imputation Station keeps them distinguishable from observed Raw Data and makes the intervention explicit. Users select variables and exclusions, review missing-code conventions, choose and configure an imputation method, and determine whether imputed values should be flagged. The current prototype provides several options, including mean/mode imputation, MICE, and random-forest-based imputation.

Figure~\ref{fig:prototype-imputation-config} shows the configuration and readiness state before execution, including conditions that can block an imputation run. Figure~\ref{fig:prototype-imputation-result} shows the resulting state, where synthesized values remain visibly marked and validation checks the intended transformation while preserving unaffected observed values. The available methods are not presented as methodological recommendations or as evidence of comparative suitability; rather, the prototype demonstrates how different imputation procedures can be embedded within the same controlled sequence of configuration, decision, intervention, validation, and documentation.

\subsection{Documentation, Replay, Final Report, and Export}
\label{sec:documentation-export}

\begin{figure}[!b]
\centering
\includegraphics[width=\columnwidth]{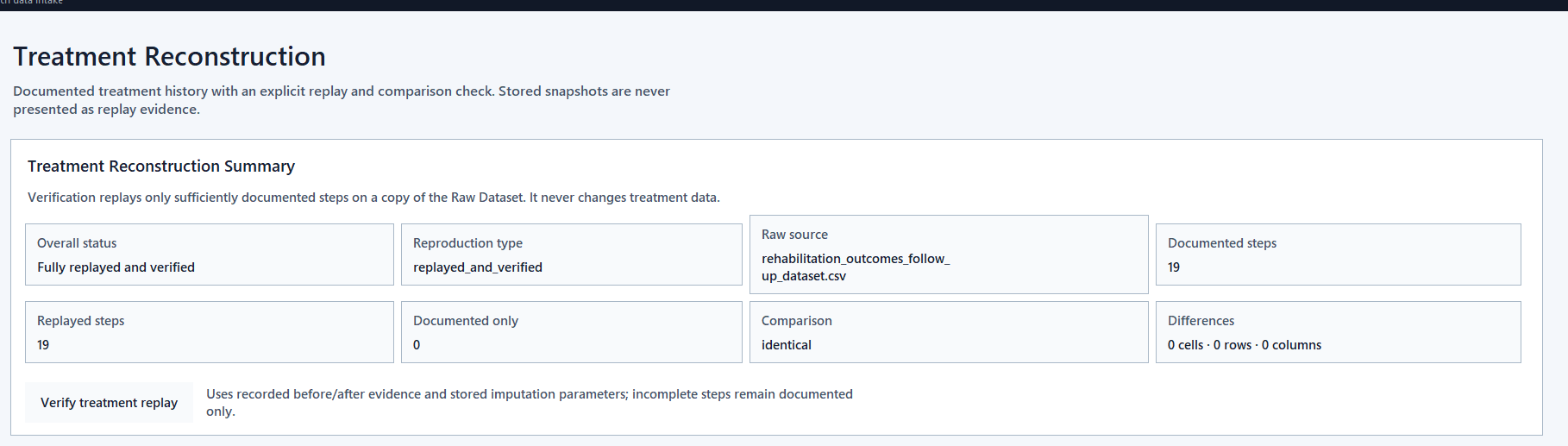}
\caption{Replay reconstruction summary and comparison with the recorded state in the current prototype.}
\label{fig:prototype-replay}
\end{figure}

\begin{figure*}[!t]
\centering
\includegraphics[width=0.96\textwidth]{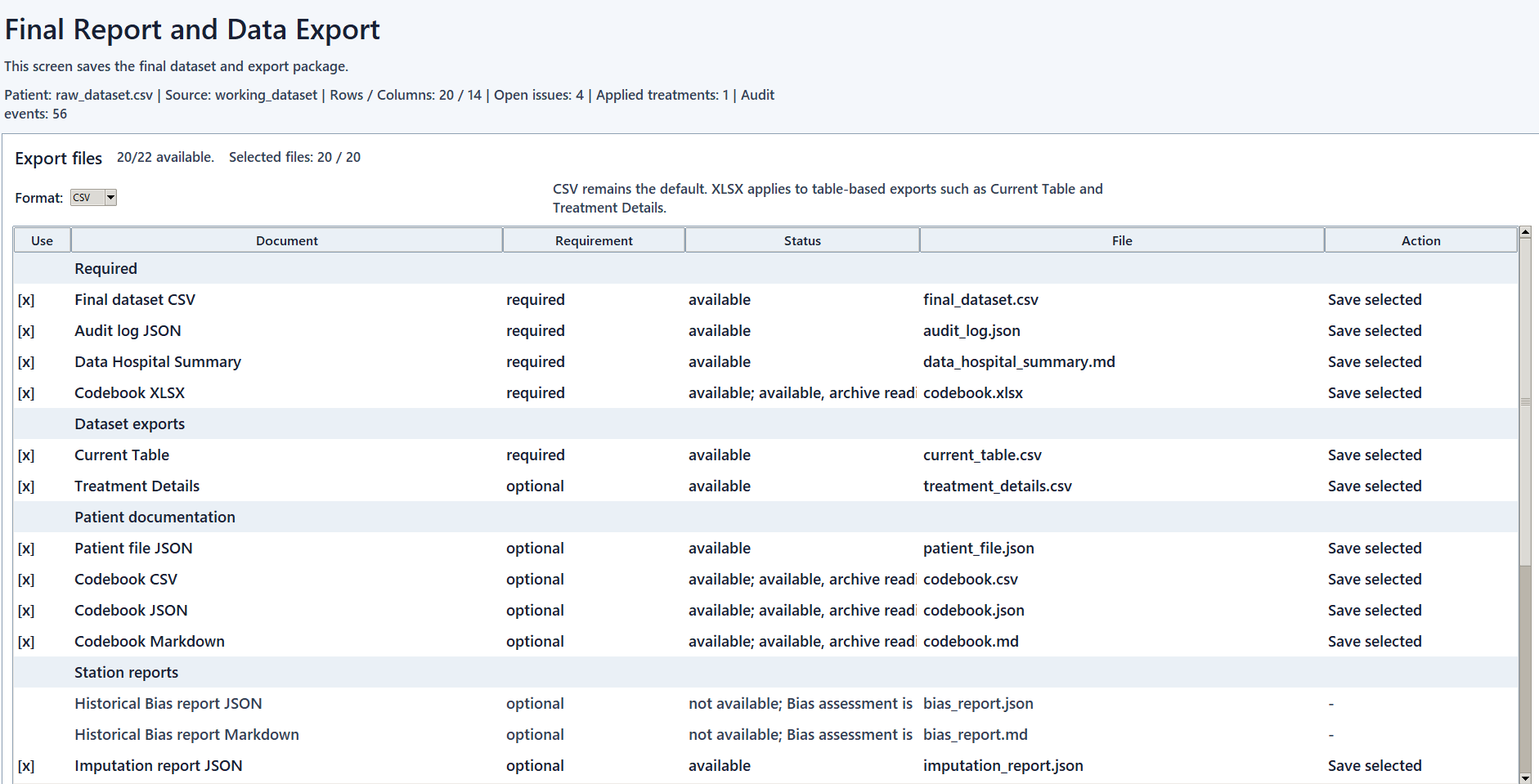}
\caption{Final Report and Data Export in the current prototype.}
\label{fig:prototype-export}
\end{figure*}

The Patient File is the growing record that connects otherwise separate interactions. It stores contextual metadata, column semantics, codebook information, open requirements, review decisions, intervention history, and provenance. The Audit Log complements this record with event-level traceability such as detected issues, evidence, proposed interventions, review decisions, affected data states, timestamps, and generated reports.

Replay extends traceability to an operational reproducibility check. Given preserved Raw Data and a sufficiently specified intervention history, the prototype reconstructs documented interventions on a copy of the original dataset and compares the reconstructed state with the recorded result. Figure~\ref{fig:prototype-replay} shows the implemented replay view. In the illustrated run, the prototype reports 19 replayed documented steps and no detected cell-, row-, or column-level differences.

Discharge leads to a Final Report and Export Package. The export view in Fig.~\ref{fig:prototype-export} exposes required and optional artifacts, including the Final Dataset, Audit Log, Data Hospital summary, codebook representations, station reports, and patient documentation. Availability is shown explicitly: outputs that do not exist in the current workflow remain unavailable instead of being presented as completed results. The export package thereby represents data quality not only as a final table, but as a documented intervention history that can be inspected and, where sufficiently specified, replayed.

\section{Planned Evaluation and Research Agenda}
\label{sec:evaluation}

Evaluation is deliberately future work: this section defines a research agenda rather than compensating for absent empirical results. Usability, explanation, decision support, and technical control address different claims and should therefore be studied separately rather than collapsed into a single overall score. Table~\ref{tab:evaluation} summarizes the planned directions.

\subsection{Evaluation Goals}
\label{sec:evaluation-goals}

Future evaluation should establish whether the \datahospital{} can support a complete data-quality task without obscuring the distinction between technical evidence, contextual judgment, user decisions, and resulting data states. It should also test whether documentation and replay remain consistent with the interventions that were actually performed. This separates evaluation of the general workflow from evaluation of any particular station-specific method.

\begin{table*}[!t]
\centering
\caption{Planned evaluation directions aligned with the Data Hospital concept}
\label{tab:evaluation}
\smalltable
\begin{tabularx}{\textwidth}{L{0.23\textwidth}Y Y}
\toprule
\textbf{Evaluation direction} & \textbf{Guiding question} & \textbf{Possible measurements} \\
\midrule
End-to-End Usability & Can users navigate the \datahospital{} and successfully intervene on identified data-quality issues without losing orientation or control? & Task completion, errors, SUS \cite{brooke1996}, NASA-TLX \cite{hart1988}, observation, interviews, and navigation traces. \\
\drdata{} / Explanation & Does evidence-bound LLM explanation improve understanding of findings, recommendations, alternatives, and uncertainty without increasing inappropriate trust? & Comprehension before/after explanations, uncertainty perception, acceptance of deliberately incorrect suggestions, qualitative justifications. \\
Decision Support / Transparency & Can users understand why a recommendation was made and make an informed, traceable decision? & Evidence-identification tasks, comparison of decisions with/without structured review, appropriateness and completeness of justifications, expert assessment. \\
Technical Validation & Does the system reliably detect, intervene on, validate, document, and reproduce known data-quality problems? & Controlled test datasets, precision/recall where applicable, state/pipeline tests, Raw Data preservation, transformation checks, imputation flags and error measures, audit consistency, replay equivalence. \\
\bottomrule
\end{tabularx}
\end{table*}

\subsection{Technical Validation}
\label{sec:technical-validation}

Technical validation should use controlled datasets for which relevant representational problems, missing-code markers, missing values, and expected intervention outcomes are known. The checks can then target the integrity of the control structure: whether Raw Data remain unchanged, approved transformations affect the intended Working Dataset, unresolved cases remain visible, imputed values are marked, validation is recomputed after intervention, audit entries correspond to performed actions, and replay reconstructs the recorded state where the intervention history is sufficiently specified.

These checks test R2, R5, R6, DP4, DP5, and DP8. Station-specific performance measures remain separate from these workflow checks. For deterministic detection or repair tasks, precision and recall can be used where a suitable reference exists. For imputation, method-specific error measures and assumptions can be evaluated in addition to the general requirements for state separation, marking, validation, and documentation.

\subsection{End-to-End Usability}
\label{sec:usability-evaluation}

A controlled user study can examine whether researchers or data stewards can move from intake and contextualization through assessment and review to an approved intervention, validation, documentation, and finalization without losing orientation or control. Task completion, interaction errors, navigation traces, observation, interviews, SUS, and NASA-TLX can provide complementary evidence about usability and workload \cite{brooke1996,hart1988}.

The representational-consistency case described in Section~\ref{sec:walkthrough-representational} provides a compact task for this purpose because the expected data transformation can be controlled while the user still has to inspect evidence and leave an ambiguous value unresolved. More consequential tasks, including imputation, can then test whether the same interaction model remains usable when assumptions and uncertainty become more important. This tests R2, R3, DP2, DP3, DP6, and DP9.

\subsection{Dr. Data and Explanation}
\label{sec:drdata-evaluation}

The explanation evaluation should test whether \drdata{} improves understanding of findings, recommendations, alternatives, and uncertainty rather than merely increasing acceptance. Suitable study designs can compare comprehension before and after an explanation, assess whether participants identify the evidence on which a recommendation is based, and examine responses to deliberately incorrect or insufficiently supported suggestions. The relevant outcome is calibrated use of the assistant, not maximal trust or maximal recommendation acceptance \cite{lee2004,amershi2019}. This tests R7, DP3, DP4, and DP7.

\subsection{Decision Support and Transparency}
\label{sec:decision-evaluation}

Decision-support evaluation should focus on whether users can explain why a recommendation was made and justify their own decision from the available evidence. Evidence-identification tasks, comparisons of decisions with and without structured review, qualitative justifications, and expert assessment can be used to evaluate transparency and decision quality. Audit records provide an additional artifact-level check because the evidence, decision, and resulting state should remain traceable after the interaction itself has ended. This tests R3, R5, DP2, DP3, and DP5.

The research agenda is deliberately staged. Controlled technical validation can first establish that the executable workflow behaves as intended on known cases. Controlled user studies can then compare a conventional cleaning interface, the structured workflow without LLM explanation, and the same workflow with \drdata{}. Such comparisons can separate effects of the control structure from effects of explanation. Exploratory field studies with real research datasets can subsequently test whether the Patient File, reports, intervention history, and replay outputs remain useful under realistic documentation and domain constraints. Concrete hypotheses, participant groups, sample sizes, datasets, and analysis plans remain decisions for those evaluation studies rather than claims of the present concept paper.

\section{Discussion}

\subsection{Opportunities and Expected Benefit}

The \datahospital{} connects technical assessment, contextual knowledge, human decision-making, controlled intervention, validation, documentation, and reproducibility. It can thereby address a gap between data cleaning tools, research data management, and explainable assistance. The combination of Patient Vitals, intervention stations, Patient File, Audit Log, and replay is particularly relevant because it models data quality as a process rather than as a single score or one-time cleaning operation.

The workflow also allows heterogeneous forms of data work to be placed under the same control logic. A representational-consistency intervention can be comparatively simple and deterministic, while imputation can require substantially more methodological assumptions and uncertainty management. Specialized methods remain specialized, but their role in the research data-quality process becomes visible and auditable through the same surrounding control structure.

The expected benefit therefore lies not only in cleaner data. The concept is intended to make relevant quality questions visible, distinguish what can and cannot be assessed, expose evidence, support informed decisions, preserve Raw Data, document interventions, prepare archiving and publication, and make sufficiently specified intervention histories replayable. In addition, the hospital metaphor may serve as a learning structure through which users explore data-quality concepts and the consequences of different decisions interactively. These are design intentions of the concept; their practical benefit remains subject to the planned evaluation in Section~\ref{sec:evaluation}.

\subsection{Risks and Limitations}

The concept has several limitations. Scores can create apparent precision if assessability is not clearly marked or if diagnostic signals are confused with established data-quality dimensions. The concept therefore separates Wang and Strong dimensions from metrics, diagnostic evidence, and workflow-specific readiness indicators. Nevertheless, the operationalization of several dimensions remains dependent on domain knowledge, reference data, or intended use and cannot be solved by interface design alone.

The current prototype and the demonstrations in this paper target tabular research data. Whether the proposed control model transfers effectively to non-tabular research data such as images, audio, spatial data, or graph-structured data has not yet been demonstrated.

LLM explanations can increase trust excessively if evidence binding and uncertainty are not visible. \drdata{} therefore remains an optional explanation layer and must not become an autonomous mutation mechanism or an unsupported source of truth. An operational deployment must also address data minimization, access control, model and prompt versioning, logging, external data transfer, and prompt injection through untrusted dataset content. The current concept establishes the control boundary but does not yet provide a complete privacy, security, or ethics architecture.

A further limitation lies in the metaphor itself. The hospital metaphor can provide orientation, but it must not suggest that data are always unambiguously sick or healthy. Many quality questions involve trade-offs between research purpose, documentation status, methodological assumptions, and risk. Likewise, a successful replay only demonstrates that a sufficiently documented intervention path can be reconstructed; it does not prove that the underlying scientific decision was appropriate.

Finally, the current prototype is non-public and therefore cannot provide independently verifiable evidence of its implementation. Future evaluation must distinguish claims about the overall control and interaction model from claims about the technical or methodological performance of individual stations.

\section{Conclusion}

The \datahospital{} is a human-in-the-loop concept for data-quality-related decisions rather than a replacement for statistical analysis, an automated cleaning system, or a FAIR certification mechanism. Its central contribution is a ten-stage control structure that connects multidimensional and purpose-dependent data quality with assessability, inspectable evidence, explicit intervention authority, controlled data states, validation, documentation, and replay.

The non-public prototype illustrates how these elements can be instantiated in an executable interface. A representational-consistency case shows evidence, optional explanation, explicit user choice, and unresolved uncertainty around a comparatively simple intervention. The Imputation Station demonstrates how the same control structure can surround a more consequential intervention that creates synthesized values, without presenting its simple baseline method as a general analytical recommendation.

The present work establishes the conceptual and methodological foundation of the \datahospital{} and aligns it with an implementation-backed demonstrator without claiming empirical effectiveness or independent reproducibility. The staged research agenda reserves technical validation, end-to-end usability, explanation, decision support, and transparency for subsequent studies. The concept paper thereby focuses on what the Data Hospital is, which control problem it addresses, and how the idea can be made concrete.

\section*{Acknowledgment}

The authors thank the members of the SynData project for valuable
discussions and feedback during the development of the Data Hospital.

This work was conducted as part of the SynData project within DataNord.
The DataNord project is funded under grant number 16DKZ2026A by the
Federal Ministry of Research, Technology and Space (BMFTR) and financed
by the European Union -- NextGenerationEU.

\balance
\bibliographystyle{IEEEtran}
\bibliography{bib/references}

\end{document}